\documentclass[aps,pra,amsmath,amssymb,superscriptaddress,reprint]{revtex4-1}

\usepackage{graphicx}
\usepackage{dcolumn}
\usepackage{bm}
\usepackage[utf8]{inputenc}
\usepackage[T1]{fontenc}
\usepackage{booktabs, array, mathptmx, float, tabularx, booktabs, lipsum, amsmath,multirow}
\usepackage{siunitx, xcolor}
\usepackage{pifont}
\usepackage[version=4]{mhchem}
\usepackage[colorlinks,linkcolor=blue,anchorcolor=blue,citecolor=blue]{hyperref}

\begin{document}


\title{Connection-oriented and Connectionless Quantum Internet Considering Quantum Repeaters}

\author{Hao Zhang}
\affiliation{Purple Mountain Laboratories, Nanjing 211111, China}

\author{Yuan Li}
\affiliation{Purple Mountain Laboratories, Nanjing 211111, China}

\author{Chen Zhang}
\email{zhangchen@pmlabs.com.cn}
\affiliation{Purple Mountain Laboratories, Nanjing 211111, China}

\author{Tao Huang}
\email{htao@bupt.edu.cn}
\affiliation{Purple Mountain Laboratories, Nanjing 211111, China}
\affiliation{State Key Laboratory of Networking and Switching Technology, Beijing University of Posts and Telecommunications, Beijing 100876, China}

\date{\today}

\begin{abstract}
With the rapid development of quantum information and technology in recent years, the construction of quantum internet for interconnecting all kinds of quantum devices, such as quantum processors and sensors, will be the next trend for practical quantum applications.
In this paper, we propose the protocols for construction of connection-oriented and connectionless quantum networks by considering the concrete quantum repeater (QR) nodes. Four classes of QRs networks are considered first and designed with two types of protocols in link layer, i.e. simultaneous and one-by-one link. Based on those two link models, the connection-oriented protocol is presented for all classes of QRs networks and the connectionless protocol is proposed for the first, second and third classes QRs networks by only one-by-one link.
Furthermore, we introduce a new hybrid connection model of quantum networks combined with connection-oriented and connectionless for practical uses.
Our work is a new attempt to study the model of the network layer for different kinds of QR networks and paves the way for developing the protocol stack of universal large-scale quantum internet.
\end{abstract}


\maketitle


\section{Introduction}
From the original quantum communication protocol Bennett-Brassard 1984 (BB84) \cite{BB84} and quantum computing model \cite{QC} to recent experimental performance of satellite-based quantum key distribution (QKD) \cite{satelliteQKD1} and quantum computational advantage \cite{Qsupremacy1,Qsupremacy2}, the rapid development of quantum information bring us the more confidence for exploring it. With the practical uses of quantum technology in the future, a large number of participants, including users and quantum devices, should be connected into a quantum internet \cite{quantumnetwork,QInternet1,QInternet2,QInternet3}. This trend can be viewed form the history of development of classical information technology over the past decades. In quantum internet, one can execute all kinds of quantum information tasks \cite{QInternet2,QInternet3,QInternet4,QInternet5}, such as securely transmitting information \cite{BB84,satelliteQKD1,E91,QSDC1,QSDCDL041,QSDCDL042,QSDCTWO1,QSDCTWO2}, realizing distributed quantum computing \cite{DQC2,DQC3}, sharing enhanced sensing signals \cite{qsensing,clocknetwork,telescopesnetwork} and etc. Therefore, capable of realizing end-to-end entanglement distribution with good quality on arbitrary two or more users is a basic and core task for benchmarking a quantum internet. However, due to the limitations of quantum technology and particularity of quantum theory, quantum internet is in its infancy and very different with its classical counterpart \cite{QInternet3}. For instance, classical communication needs repeaters to extend communication distance because of channel loss, and the quantum channel also requires quantum repeater (QR) to overcome its loss and decoherence but with absolutely different principle due to the non-clone theorem of unknown quantum state \cite{QRBDCZ,QRRMP}.

To build a practical quantum internet, the quantum technologies that meets the requirements, like good quantum memories \cite{QRRMP,QRDLCZ,QRHOM,QRwang,QRencoding1,QRencoding2,QRencoding3}, fast quantum gates operations \cite{QRQEC1,QRQEC2,QRQEC3} or special entangled state generations \cite{QRAP}, are the basic elements that should be improved \cite{QRreview}. This is because the good quantum technologies can promise the feasibility and stability of physical layer. Besides, constructing an universal quantum internet protocol stack, which is in analogy with the operating system of a computer, is another core architecture in terms of the network itself \cite{QInternet2}. The execution of quantum operations are controlled by a series of classical signals at the present stage. Therefore, assisted with classical internet, the design of quantum internet protocol stack becomes intriguing and more complicated. In recent years, some meaningful researches about quantum internet are appeared with range from link layer \cite{Qinternetlink1,Qinternetlink2,PacketS} to network layer \cite{Erouting1,Erouting2,Erouting3,Erouting4,Qinternetexp,CircuitS} and even to whole protocol stack \cite{quantumstack1,quantumstack2,quantumstack3}. To promote the quantum internet into mature in the future, there are still many studies to do for improving fundamental quantum technologies and quantum internet system as well as building the bridge between them.

In this paper, we propose the connection-oriented and connectionless protocols for end-to-end entanglement distribution in large-scale quantum network in terms of the different kinds of QR nodes. Four classes of QRs are divided into two link models according to their architectures. Results show that the first, second and all-photonic QRs networks can be designed with simultaneous link protocol for building the connection-oriented quantum network. Another one-by-one link protocol can be applied for both the connection-oriented and connectionless quantum networks with the candidates of the first, second and third classes QRs networks. Furthermore, we combine the connection-oriented and connectionless protocols to create a hybrid network connection model to show their flexibility in practical uses. All the protocols are given with detailed steps.
Our work paves the way to construct the network layer of practical quantum internet by considering different QR nodes with the help of the development history of classical internet.

The article is organized as follows: In Sec.\ref{secQI}, we introduce the entanglement-based quantum internet and four classes of QRs.
In Sec. \ref{secEDNL}, three types of quantum network protocols, i.e. connection-oriented, connectionless and hybrid connection, are presented based on different link protocols.
In Sec. \ref{secDP} and \ref{secsummary}, a discussion and summary are given, respectively.

\begin{figure}[t]
\includegraphics[width=\linewidth]{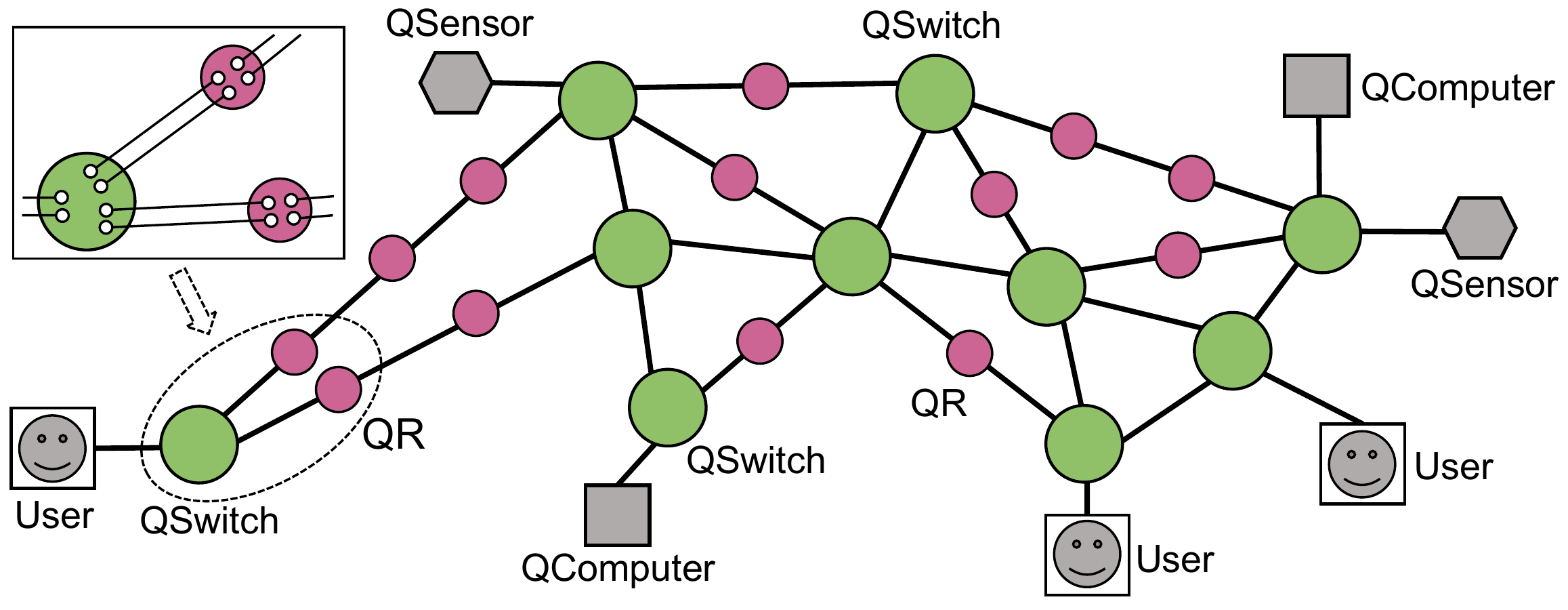}
\caption{Schematic diagram of large-scale quantum network with many nodes. QComputer, quantum computer; QSensor, quantum sensor; QR, quantum repeater; QSwitch, quantum switch with functions of both quantum router and repeater. Some quantum devices are not shown, such as quantum interfaces, entanglement sources and etc. All the nodes are connected by physical channel and assisted by classical internet. Insert is the partial enlarged schematic diagram of dashed ellipse. The white circles represent qubits and the black lines are physical channels which can be optical fibres or free space. The practical uantum networks are more complex and the amount of qubits in quantum nodes are usually more than the presentation in the figure.}\label{qinternet}
\end{figure}

\section{Entanglement-based Quantum Internet}\label{secQI}

\subsection{Large-Scale Quantum Networks}\label{secQNTQN}
As the distance and the number of users increase, the connections between many quantum nodes form a large-scale quantum network on which quantum tasks can be executed, like quantum communications \cite{BB84,satelliteQKD1,E91,QSDC1,QSDCDL041,QSDCDL042,QSDCTWO1,QSDCTWO2}, quantum computations \cite{Qsupremacy1,Qsupremacy2,DQC2,DQC3,BQC} and quantum sensing \cite{qsensing,clocknetwork,telescopesnetwork}. Actually, today's quantum information tasks can be divided into two classes based on the use of quantum resource for building end-to-end quantum channel. One is single photon channel used for protocols like BB84 QKD \cite{BB84} and Deng-Long04 quantum secure direct communication (QSDC) \cite{QSDCDL041,QSDCDL042}, and another one is entanglement whose applications include Ekert91 QKD \cite{E91}, two-step QSDC \cite{QSDCTWO1,QSDCTWO2}, distributed quantum computing \cite{DQC2,DQC3} and two/three-server blind quantum computing \cite{BQC}. In practical experiments, single photon channel is easier than entanglement one, but in this article we mainly consider the entanglement network for two reasons. One reason is that all the single photon protocols can be replaced by entanglement version theoretically. For instance, one can use E91 protocol or quantum teleportation \cite{teleportation} to complete quantum key agreement of BB84, however in distributed quantum computing or sensing which takes advantage of nonlocal quantum correlation, the role of entanglement can not be replaced by single photon. Another reason is the early QR models are based on entanglement distribution in long distance quantum channel. A schematic diagram of quantum network is shown in Fig. \ref{qinternet}. With the help of classical internet, the end-to-end entanglement distribution can be completed between users and quantum servers by using QRs and quantum switches. The quantum switch can be a multi-port QR considered in the insert of Fig. \ref{qinternet} or a pure quantum router. To construct a quantum channel, the core task is to distribute the one of Bell states between two nodes as follows,
\begin{eqnarray}\label{Bell}
|\Psi_{\pm}\rangle&=&\frac{1}{\sqrt{2}}(|00\rangle\pm|11\rangle),\\
|\Phi_{\pm}\rangle&=&\frac{1}{\sqrt{2}}(|01\rangle\pm|10\rangle).
\end{eqnarray}
With above end-to-end entanglement channel, quantum information tasks can be executed with corresponding quantum protocols. For instance, users can securely communicate with other users, or delegate their complex computing tasks using blind quantum computing protocol on servers who has quantum computers. Indeed, some protocols of quantum communication, such as BB84, also can be performed using single photon in quantum network shown in Fig. \ref{qinternet}. In this case, the long distance quantum state transmission can make use of QR which is based on error correction in principle.

\begin{figure}[t]
\includegraphics[width=\linewidth]{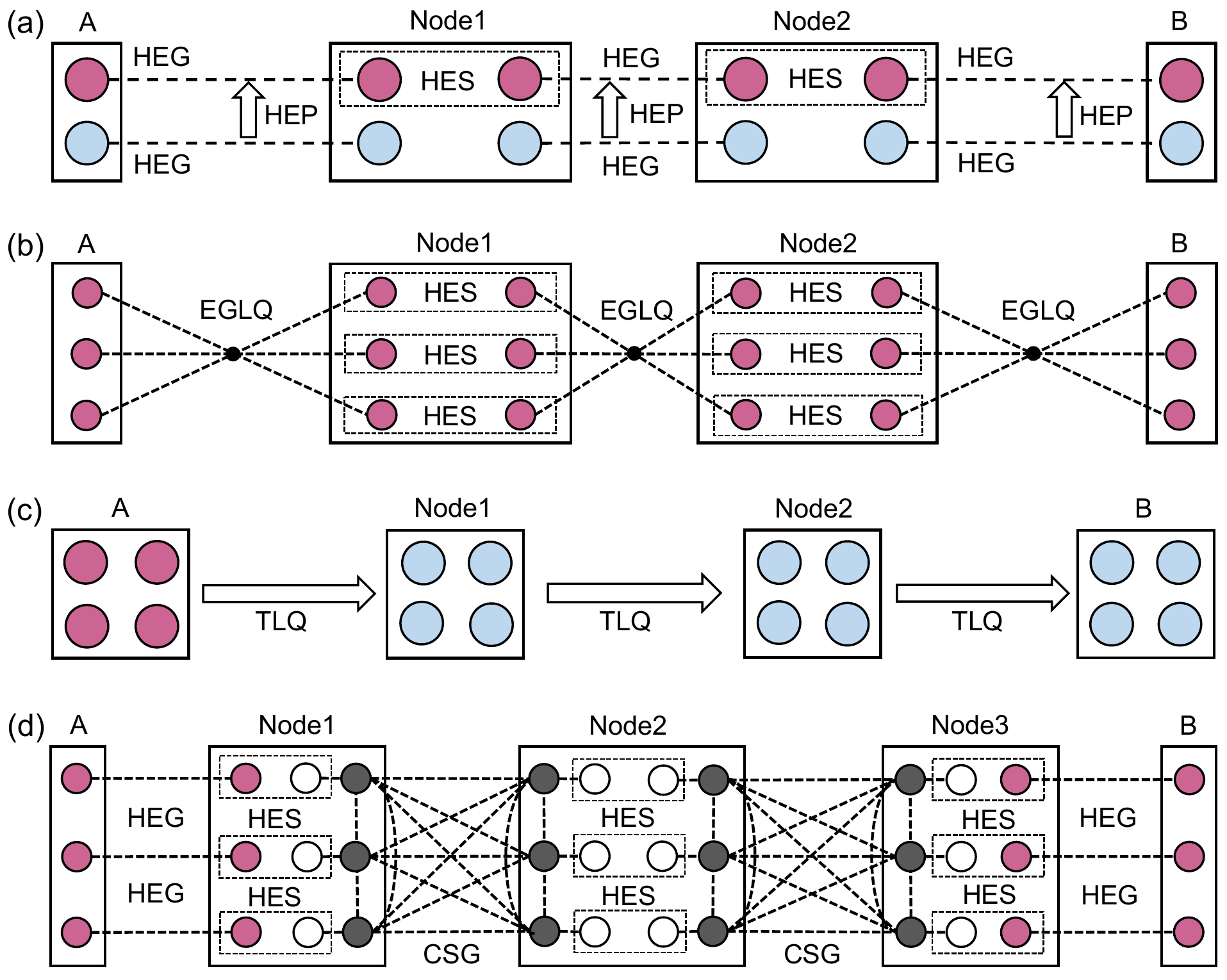}
\caption{Major architectures of four classes of QRs. HEG, heralded entanglement generation; HEP, heralded entanglement purification; HES, heralded entanglement swapping; EGLQ, entanglement generation of logical qubits; TLQ, transmission of logical qubits; CSG, cluster state generation. (a) The first class QR with HEG, HEP and HES \cite{QRBDCZ,QRDLCZ}. (b) The second class QR with EGLQ and HES \cite{QRencoding2,QRencoding3}. (c) The third class QR with TLQ \cite{QRQEC2,QRQEC3}. The logical qubits are encoded with many physical qubits. The teleportation-based error correction is performed in each node. (d) All-photonic QR with HEG, CSG and HES \cite{QRAP}.}\label{qrepeater}
\end{figure}

\subsection{Quantum Repeaters}\label{secQR}
For building long distance quantum channel, QR is an indispensable quantum device, and sometimes it acts as the quantum switches with multi-port used for function of both repeaters and routers in quantum networks. The first QR model is proposed by Briegel et al. in 1998 \cite{QRBDCZ}. Nowadays, the existing QRs are mainly divided into four classes, i.e. the first to third class \cite{QRBDCZ,QRRMP,QRDLCZ,QRHOM,QRwang,QRencoding1,QRencoding2,QRencoding3,QRQEC1,QRQEC2,QRQEC3} and all-photonic QRs \cite{QRAP}, as shown in Fig. \ref{qrepeater}. The first class QR in Fig. \ref{qrepeater} (a) uses heralded entanglement generation and purification \cite{HEP1,HEP2,HEP3,HEP4,HEP5,HEP6,HEP7,HEP8,HEP9,HEP10} for correcting loss and operation errors  \cite{QRBDCZ,QRDLCZ,QRHOM}, respectively. The next step is entanglement swapping completed for extending the entanglement distance. In entanglement generation, purification and swapping, the good quantum memories are required in each quantum nodes for keeping the coherence of quantum states \cite{QM1,QM2,QM3,QM4}. The representative scheme is Duan-Lukin-Cirac-Zoller (DLCZ) model which utilizes the Raman process of three-level atomic ensemble \cite{QRDLCZ}. The second class QR in Fig. \ref{qrepeater} (b) is designed with encoding for eliminating operation errors \cite{QRencoding1,QRencoding2,QRencoding3}, such as Calderbank-Shor-Steane encoding \cite{QRencoding2}, and still with entanglement generation and swapping for overcoming photon loss. Different with above two classes of QRs, the third class is realized with complete error correction code \cite{QRQEC1,QRQEC2,QRQEC3} shown in Fig. \ref{qrepeater} (c), so the quantum state can be transferred with logical qubits between nodes directly without entanglement generation,  purification and swapping except for surface code QR \cite{QRQEC1} which needs the entanglement channel for connecting the physical qubits in each quantum nodes. The third class protocols have lower requirement on quantum memories but need the fast quantum gate operations and large physical qubits.
Besides in Fig. \ref{qrepeater} (d), the architecture of all-photonic QRs which uses cluster states as the quantum resource for connections between adjacent nodes is similar with the first and second class but without needing good memories in some cases \cite{QRAP}.
According to the practical experimental conditions, such as coupling efficiency, speed and fidelity of quantum gates and etc \cite{QRreview}, one can choose the corresponding QR nodes for constructing quantum networks.

\begin{figure*}[]
\includegraphics[width=\linewidth]{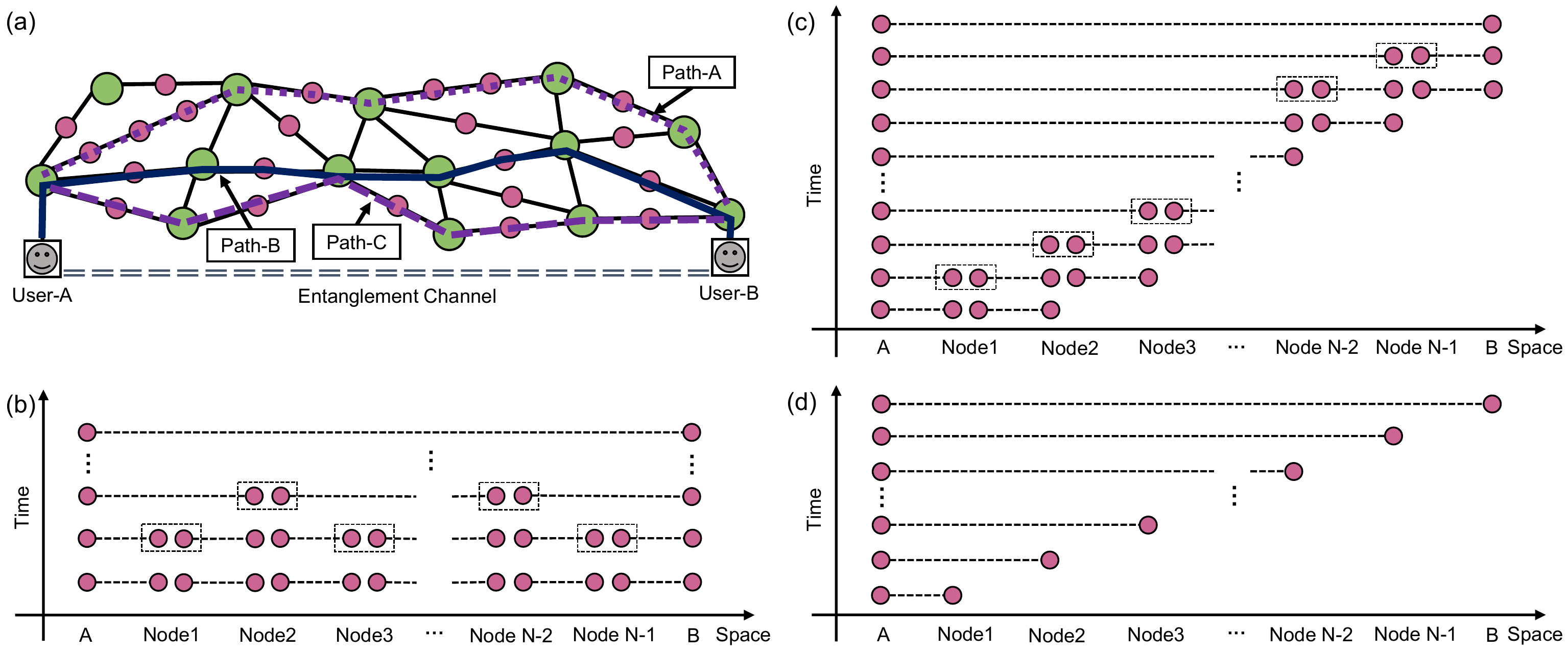}
\caption{(a) Construction of end-to-end entanglement channel in connection-oriented quantum internet. The path-B is chosen for performing the task by classical internet system. All quantum nodes are bundled with corresponding classical nodes. (b) Simultaneous link protocol of the network of the first, second class and all photonic QRs. (c) One-by-one link protocol for networks of the first and second class QRs nodes. Entanglement generation, purification and error corrections are used but not shown. (d) One-by-one link protocol for the third class QRs networks. Error corrections are used but not shown.}\label{qrconnecto}
\end{figure*}

\section{End-to-end Entanglement Distribution in Network Layer}\label{secEDNL}

According to the transmission protocol of communication channel in network layer, the classical network has two models, so called connection-oriented and connectionless. In connection-oriented networks, a fixed path is chosen first by system's algorithm and then users transmit the signals along this path to the destination. The typical example is early telephone communication using circuit switch protocol. This type connection has the advantage of keeping the good and stable quality of communications. Different with connection-oriented networks, the path of communication channel is not fixed in advance in connectionless networks. In classical network, the path is determined by the header of packet compiled with address based on the routing table of whole network system. It is suitable for network with large number of business requirements for its high extent of using network resource.
Very different from classical networks, we consider the connection-oriented and connectionless models only for the process of constructing the quantum channel in quantum entanglement networks rather than executing the high-level protocols in completed entanglement channel. Because after the entanglement channel is constructed successfully, the performance of high level protocols has no relation with the previous physical path and the nodes in quantum network except for end nodes in terms of task with only one round.
In this work, we assume that when the path is reserved in quantum networks, the enough quantum resources, including channel capacity, quantum memory qubits and etc, are prepared for building quantum channels.

\subsection{Connection-oriented Model}\label{secCOM}
In practical large-scale quantum networks, how to distribute end-to-end entanglement with specific business requirements belongs to a problem of network layer. Here we propose a connection-oriented protocol of quantum internet by considering the different kinds of QRs. As shown in Fig. \ref{qrconnecto} (a), we assume that the user-A want to building a quantum channel with user-B. Considering the connection-oriented model, user-A should first send request to the classical internet. Subsequently, the classical central control system or the distributed routers will choose an appropriate path, such as choosing path-B, for them to construct the connection based on preset algorithms. At this point, the classical internet system coordinates all the nodes of this determined path to distribute entangled states, such as Eq. (\ref{Bell}), according to the type of QRs and quantum switches.

Considering the networks of four classes of QRs, we first introduce two types of entanglement distribution protocols of link layer shown in Fig. \ref{qrconnecto} (b)-(d), i.e. simultaneous link and one-by-one link. Fig. \ref{qrconnecto} (b) is a simultaneous link protocol which needs first make connections between adjacent nodes simultaneously. Subsequently, the entanglement purifications and swapping or error corrections are performed to extended the entanglement channels. It is suitable for the first, second class and all-photonic QRs in principle. However, the third class QR networks can not uses simultaneous link. Different with simultaneous link, the one-by-one link protocol distributes the entanglement from one node to the neighbor one and subsequently extends the entanglement to the end node by hopping all the quantum nodes in its path one by one. The quantum networks with the first, second and third classes QRs can use this link protocol. For the first and second classes QRs networks, the process of one-by-one link is designed in Fig. \ref{qrconnecto} (c), and the protocol for the third class QR is shown in Fig. \ref{qrconnecto} (d). One-by-one link is not considered for the all-photonic QRs nodes. In Fig. \ref{qrconnecto} (c) and (d), each hop is executed by doing the entanglement purifications for the first class QR, but for the second and third classes, the hopping operations are only error corrections. In some cases, such as short distance networks or space satellite quantum networks \cite{spaceq}, the link protocol in Fig. \ref{qrconnecto} (d) also can be executed using only purifications and error corrections for the first and second class QRs nodes with low photon loss.

For connection-oriented quantum networks, both the simultaneous and one-by-one link protocol can be used for building the entanglement channel. However, for the first and second classes QR nodes, the simultaneous link protocol is better as the built-up time of one-by-one protocol is longer.
The longer time means the more decoherence of qubits in quantum memories and longer delay. If considering the one-by-one link protocol, the third class QR network is more appropriate.
The detailed steps of connection-oriented quantum internet shown in Fig. \ref{qrconnecto} (a) are roughly given as follows,

Step 1, User-A sends the request to classical internet system. The request includes information about the users' address and other requirements, such as class of QR, protocol of link layer, i.e. simultaneous or one-by-one link, time delay and etc.

Step 2, When the classical internet system receives the request signal, it will calculate a path based on the preset algorithm according to the user's requirements, topology and traffic condition of quantum network. For instance, the path-B is chosen from some candidates in Fig. \ref{qrconnecto} (a). This process can be done with different classical network operation mechanisms, such as classical central control system or distributed routers model. Then the operational orders will be sent to each quantum node in this chosen path via classical channels.

Step 3, All the quantum nodes in the path receive the operational orders and construct the entanglement channel based on the class of QRs nodes with simultaneous link in Fig. \ref{qrconnecto} (b) or one-by-one link protocols in Fig. \ref{qrconnecto} (c) and (d). For the first class QR network, this process includes heralded entanglement generations, purifications, swapping, and for the second class, the purifications are replaced by error corrections. The third class QR networks are only operated with error corrections. For the all-photonic QR nodes, the cluster states should be prepared at the first step of simultaneous link protocol.

Step 4, The end-to-end entanglement channel is completed and the quantum nodes, except for two end nodes, in the path are free for other uses. The successful confirmation between two end nodes are triggered by the last heralded signals from the middle quantum node for simultaneous link or the signal from user-B for one-by-one link protocol.

Just like the classical internet, the locations of all the nodes and the topology of whole quantum network are known and labelled with numbers for calculating the path in connection-oriented quantum internet.
In the process of distributing entanglement, one can choose both the circuit and packet switching protocols \cite{PacketS} of quantum version with different service requirements. For example, the circuit switching protocol is preferred for service customized quantum network which pursues high quality, and those networks are usually the time sensitive networks used for industrial quantum internet, medicinal and military applications.

\subsection{Connectionless Model}\label{secCM}
\begin{figure}[]
\includegraphics[width=\linewidth]{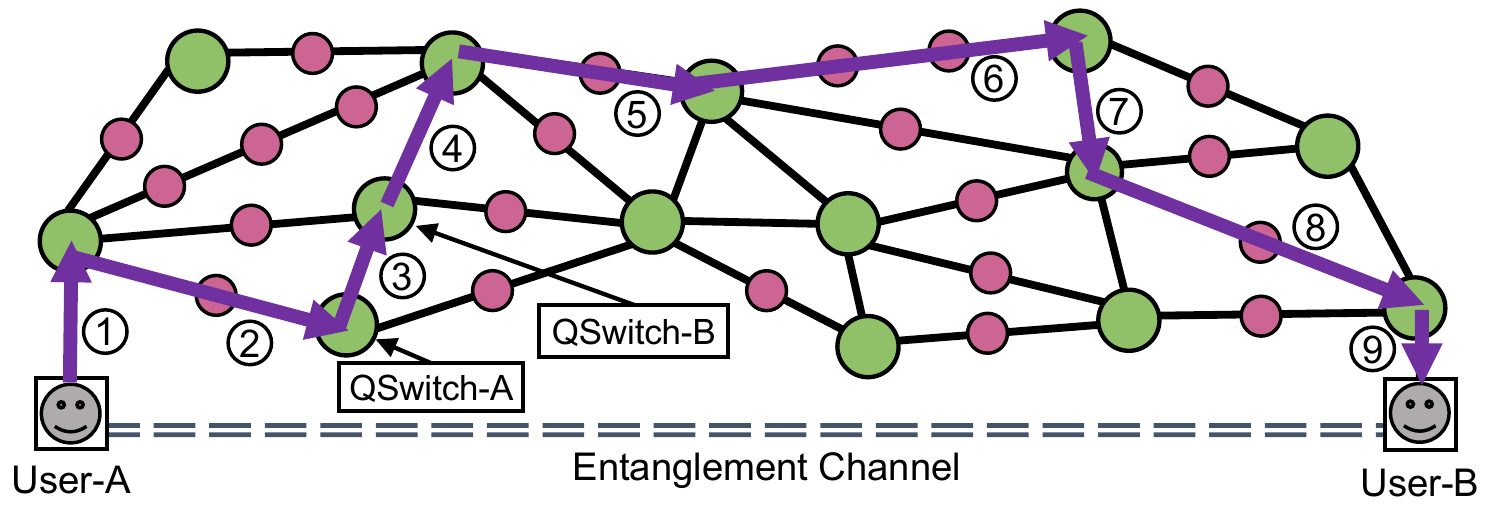}
\caption{Construction of end-to-end entanglement channel in connectionless quantum internet. The path labelled with number 1-9 is chosen one after another by using packet switching technique.}\label{qrconnectl}
\end{figure}

In addition to connection-oriented quantum internet, it has another model of connectionless.
Connectionless quantum networks shown in Fig. \ref{qrconnectl} can be realized by using only one-by-one link shown in Fig. \ref{qrconnecto} (c) and (d) with packet switching technique. The first, second and third classes QR networks can use this protocol. However, because the simultaneous link protocol is not available in connectionless, it indicates circuit switching protocol can not work here. The connectionless model is not considered with the all-photonic QRs nodes.

The packet switching protocol of quantum signal is introduced in reference \cite{PacketS}. The quantum signal is inserted as a quantum payload in a quantum frame structure between a classical header and trailer. The classical header is designed for carrying information used for routing, error corrections, entanglement purifications and swapping, and the trailer indicates the end of the quantum frame structure. Therefore, the quantum signal can be not only single photon but a large number of photons used for purifications or error corrections in long distance quantum channel. With the help of the classical header, the quantum signal can arrive the destination successfully.
As shown in Fig. \ref{qrconnectl}, the main steps of connectionless quantum networks are given as follows,

Step 1, User-A compiles user-B's address in a classical header and arranges the half of photons of entangled pair as a quantum payload by using packet switching technique. Then the quantum frame structure composed of classical header, quantum payload and classical trailer is sent out.

Step 2, When the quantum frame structure arrives the first quantum switch, it will be routed to the next quantum channel according to the classical header. Different with connection-oriented model, the path here is not chosen before transmission and it is decided by present router based on the routing table. Therefore, the next channel is arranged in current node. As shown in Fig. \ref{qrconnectl}, as an example, the signals in a path is routed from channel 1 to 9.

Step 3, The quantum nodes receive the quantum frame structure and extend the entanglement channel based on one-by-one link. For the first and second classes QR networks, when the next channel is chosen, example given with path-3 shown in Fig. \ref{qrconnectl} which is determined while classical header arrives the quantum switch-A, the entanglement channel in channel-3 should be constructed immediately for the purpose of extending the entanglement distance. This heralded entanglement generation in channel-3 can be executed at the same time as the purifications or error corrections of channel-2 for saving the time. In both QR and quantum switch nodes, operations including the entanglement purifications or error corrections should be done. When the classical header arrives the switch-B, the quantum payload should be replaced with new entangled photons hold by switch-B. When the quantum signal arrives the end node, the entanglement channel is constructed after the entanglement swapping. For the third class QR network, the quantum payload is corrected by error correction operations in each QR nodes. In the end node, the error corrections or entanglement purifications are also indispensable to improve the fidelity. The quantum node is free after the entanglement swapping for the first and second class QR networks. In terms of the third class QR networks, the quantum node is released while the quantum frame structure leaves.

Step 4, User-B confirms the successful arrival of quantum payload with user-A. If the user-A does not receive the confirmation signaling within the predetermined time from user-B, user-A defaults that task is failure and repeats the above procedures.

For connectionless model, the path in next round of constructing the quantum channel will be probably different with previous one. The classical central control system also should be used in connectionless for some special uses. For instance, one want to control the traffic flow of some specified paths. In classical internet, this connectionless model is so called ``best-effort''.

\begin{figure}[]
\includegraphics[width=\linewidth]{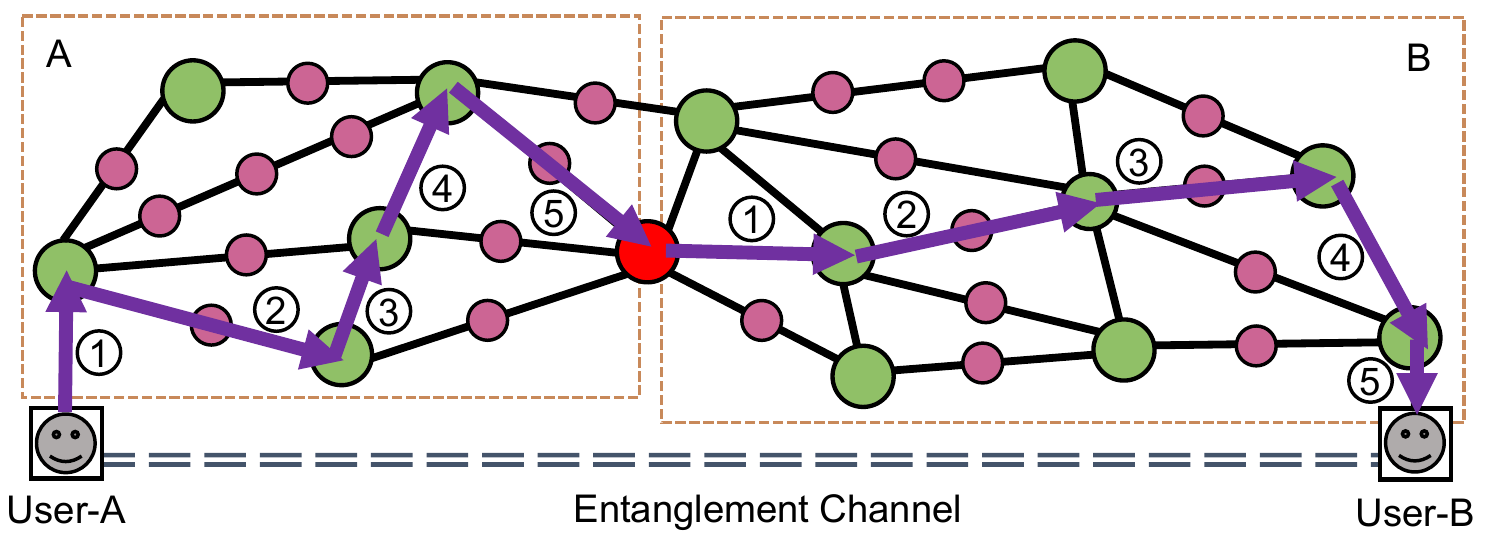}
\caption{Hybrid connection model for construction of end-to-end entanglement channel in quantum internet. The red quantum node is the fixed path of connection-oriented protocol. A and B are two connectionless areas connected by red node.}\label{qrconnecth}
\end{figure}

\begin{table*}[]
\caption{Characteristics of four classes of QRs nodes in quantum networks. HEG, heralded entanglement generation; HEP, heralded entanglement purification; HES, heralded entanglement swapping; ECC, error correction code; GQM, good quantum memory; FGO, fast gate operation; SL, simultaneous link; OL, one-by-one link; CO, connection-oriented; CL, connectionless. \ding{51}(\ding{55}) represents need (no need); $\square$, to be selected; NC, not under consideration. }\label{fourqr}
\begin{ruledtabular}
\begin{tabular}{cccccccccccc}
 QRs & HEG & HEP & HES & ECC & GQM & FGO & SL & OL & CO & CL \\
\hline
First-Class& \ding{51} & \ding{51} & \ding{51} & \ding{55} & \ding{51} & \ding{55} & \ding{51} & \ding{51} & \ding{51} & \ding{51} \\
Second-Class\footnotemark[1]& \ding{51} & \ding{55} & \ding{51} & \ding{51} & \ding{51} & \ding{55} & \ding{51} & \ding{51} & \ding{51} & \ding{51} \\
Third-Class& \ding{55} & \ding{55} & \ding{55} & \ding{51} & \ding{55} & \ding{51} & \ding{55} & \ding{51} & \ding{51} & \ding{51}  \\
All-Photonic\footnotemark[2] & \ding{51} & $\square$ & \ding{51} & $\square$ & \ding{55} & $\square$ & \ding{51} & NC & \ding{51} & NC  \\
\end{tabular}
\end{ruledtabular}
\footnotetext[1]{For the second class QRs, the requirement of GQM is lower than the first class QRs due to the use of ECC to replace HEP. GOM can reduce the requirement of FGO, and FGO can reduce memory time in turn.}
\footnotetext[2]{For the all-photonic QRs, HEG includes generation of Bell and cluster states. HEP and ECC are chosen according to the concrete conditions. GQM is unnecessary in principle but also required with special cases.}
\end{table*}

\subsection{Hybrid Connection Model}\label{secHCM}
The connection-oriented and connectionless model have their own advantages. Sometimes, the hybrid model which combines the connection-oriented and connectionless protocol has its new practical applications. We show an example of the hybrid model in Fig. \ref{qrconnecth}. The red quantum node and two end nodes chosen with a fixed points are connection-oriented. The red node divides the whole quantum network into two areas A and B. In this two areas, the connectionless protocol is applied. In large-scale quantum networks, more areas with connectionless can be divided.
The steps of protocol are given as follows,

Step 1, User-A sends the request to classical internet system with information about the users' address and other requirements.

Step 2, The classical internet system will choose a connection-oriented nodes, such as red node and two end nodes in Fig. \ref{qrconnecth}, and sends the operational orders to those nodes.

Step 3, In area A and B, the connectionless entanglement distribution can be executed as steps 1-4 introduced in Sec. \ref{secCM} to extend the channel along 1-2-3-4-5 in area A and B of Fig. \ref{qrconnecth}. At this point, the entanglement channel is built between user-A and red node, and the same for user-B and red node.

Step 4, Entanglement swapping is done for completing the final entanglement channel between user-A and B. And the last heralded signals are sent to the user-A and B to confirm the success.

The version of above step 3 is a fast scheme. One can also choose pure one-by-one link protocol form user-A to B but with the condition that the red node that must be passed. This connection in link layer will cost more time in long distance entanglement distribution.

\section{Discussion}\label{secDP}

In connection-oriented quantum internet, the classical central control system which used for control the operations of all quantum nodes can be realized with software defined network \cite{SDN} which supervises all the classical nodes in network via only one or a few "brain".  Compared with the classical internet, the operations in quantum internet are more complex and needs more precise control. For instance, the heralded entanglement generation and swapping in first class QRs, such as DLCZ model, is sensitive to the synchronization of adjacent quantum nodes. Therefore, a perfect and powerful central control system is crucial in this model.

For connectionless quantum internet, the hybrid packet switching protocol is used and the routing process is determined by classical header which uses the classical network. With the address information in classical header, the quantum signal will be improved in its fidelity when it passes each quantum node, and finally arrive the end node. The path chosen is completed via classical routers and the quantum signals are dealt with quantum nodes. This requires the good cooperativity of local classical and quantum devices.

It is similar to the classical network in network perspective, due to the reservation of network resource, connection-oriented quantum internet has advantage of keeping good quality for construction of quantum channel and can be further used for deterministic and service customized quantum network. But this network model can not make use of network resource with maximum extent while there are so many tasks in quantum networks. However in connectionless, it holds the advantage in utilization of network resource but without deterministic service quality due to the loss of quantum frame structure in the process of transmission with limited network resources in both classical and quantum internet. Therefore, the hybrid connection protocol which combines the characteristics of connection-oriented and connectionless is a tradeoff choice between quality and utilization.

For quantum communications based on single photon, one can use the quantum teleportation to transfer the quantum state. Besides, in the third class QR network, the single photon can be transmitted directly with encoding in both connection-oriented and connectionless protocols.

In practical large-scale quantum internet, many good quantum technologies are required, such as very good memories, fast quantum gate operations, various quantum interfaces and etc. Good quantum memory plays an fundamental and important role in quantum information, as this is not only for supporting QRs but also storing the quantum states which needs long time keeping \cite{QM1,QM2,QM3,QM4}. In QR, quantum memory is used for entanglement purifications for improving the fidelity of quantum state \cite{HEP1,HEP2,HEP3,HEP4,HEP5,HEP6,HEP7,HEP8,HEP9,HEP10}. Fast quantum gate operation is indispensable for the quantum error correction which is a basic technique for overcoming decoherence in both fault-tolerant quantum computing and communication. To show the results clearly, we summarize the characteristics of four classes of QRs networks in Tab. \ref{fourqr}. In large quantum networks, converting the frequency between the telecom wavelength and microwave and transferring the quantum state between photon and atom are common operations for connecting the quantum communication and computing, therefore they need the quantum interfaces \cite{Qinterface1,Qinterface2}.

\section{Summary}\label{secsummary}
To build a bridge between the fundamental quantum devices and quantum internet system, in this paper, we propose the connection-oriented and connectionless protocol of quantum internet by considering the QRs. Four classes of QRs are considered for constructing the link layer, i.e. simultaneous and one-by-one link protocols and further used for connection-oriented and connectionless model in network layer. In terms of link layer, the third class QRs networks are not suitable for the simultaneous and the all-photonic QRs are not considered in the one-by-one link protocols. For network layer, both the simultaneous and one-by-one link protocols can be applied for connection-oriented quantum network, but only the one-by-one link is suitable for connectionless. At last, we extend the connection-oriented and connectionless model to a hybrid connection protocol.



\end{document}